\documentclass[11pt,a4paper]{article}
\usepackage{lmodern}
\usepackage[dvipdfmx]{graphicx}
\usepackage{float}
\usepackage{amsmath}
\usepackage{multicol}
\usepackage{lipsum}
\usepackage[center,small]{caption}
\usepackage{color}
\usepackage{epsf}
\usepackage{epsfig}
\usepackage{url}
\usepackage{braket}
 \usepackage[compress]{cite}
 \usepackage{amsmath,amssymb}
 \usepackage{enumitem}

\usepackage{subfig}
\usepackage{lineno}

\newcommand{\dif}{\mathrm{d}}

\begin{document}
\begin{tabular*}{15.cm}{l@{\extracolsep{\fill}}r}
{Published in JCAP as DOI: 10.1088/1475-7516/2017/06/026} 
\end{tabular*}

\vspace*{0.5cm}
\begin{center}
{\large
{\bf
Multi-resolution anisotropy studies of ultrahigh-energy cosmic rays detected at the Pierre Auger Observatory
}}\\
\vspace*{.5cm}
\normalsize {

{\bf The Pierre Auger Collaboration}

A.~Aab$^{63}$,
P.~Abreu$^{70}$,
M.~Aglietta$^{48,47}$,
I.~Al Samarai$^{29}$,
I.F.M.~Albuquerque$^{16}$,
I.~Allekotte$^{1}$,
A.~Almela$^{8,11}$,
J.~Alvarez Castillo$^{62}$,
J.~Alvarez-Mu\~niz$^{79}$,
G.A.~Anastasi$^{38}$
% ...

\begin{description}[labelsep=0.2em,align=right,labelwidth=0.7em,labelindent=0em,leftmargin=2em,noitemsep]
\item[$^{1}$] Centro At\'omico Bariloche and Instituto Balseiro (CNEA-UNCuyo-CONICET), Argentina
\item[$^{2}$] Centro de Investigaciones en L\'aseres y Aplicaciones, CITEDEF and CONICET, Argentina
\item[$^{3}$] Departamento de F\'\i{}sica and Departamento de Ciencias de la Atm\'osfera y los Oc\'eanos, FCEyN, Universidad de Buenos Aires, Argentina
\item[$^{4}$] IFLP, Universidad Nacional de La Plata and CONICET, Argentina
\item[$^{5}$] Instituto de Astronom\'\i{}a y F\'\i{}sica del Espacio (IAFE, CONICET-UBA), Argentina
\item[$^{6}$] Instituto de F\'\i{}sica de Rosario (IFIR) -- CONICET/U.N.R.\ and Facultad de Ciencias Bioqu\'\i{}micas y Farmac\'euticas U.N.R., Argentina
\item[$^{7}$] Instituto de Tecnolog\'\i{}as en Detecci\'on y Astropart\'\i{}culas (CNEA, CONICET, UNSAM) and Universidad Tecnol\'ogica Nacional -- Facultad Regional Mendoza (CONICET/CNEA), Argentina
\item[$^{8}$] Instituto de Tecnolog\'\i{}as en Detecci\'on y Astropart\'\i{}culas (CNEA, CONICET, UNSAM), Centro At\'omico Constituyentes, Comisi\'on Nacional de Energ\'\i{}a At\'omica, Argentina
\item[$^{9}$] Observatorio Pierre Auger, Argentina
\item[$^{10}$] Observatorio Pierre Auger and Comisi\'on Nacional de Energ\'\i{}a At\'omica, Argentina
\item[$^{11}$] Universidad Tecnol\'ogica Nacional -- Facultad Regional Buenos Aires, Argentina
\item[$^{12}$] University of Adelaide, Australia
\item[$^{13}$] Centro Brasileiro de Pesquisas Fisicas (CBPF), Brazil
\item[$^{14}$] Universidade de S\~ao Paulo, Escola de Engenharia de Lorena, Brazil
\item[$^{15}$] Universidade de S\~ao Paulo, Inst.\ de F\'\i{}sica de S\~ao Carlos, S\~ao Carlos, Brazil
\item[$^{16}$] Universidade de S\~ao Paulo, Inst.\ de F\'\i{}sica, S\~ao Paulo, Brazil
\item[$^{17}$] Universidade Estadual de Campinas (UNICAMP), Brazil
\item[$^{18}$] Universidade Estadual de Feira de Santana (UEFS), Brazil
\item[$^{19}$] Universidade Federal de Pelotas, Brazil
\item[$^{20}$] Universidade Federal do ABC (UFABC), Brazil
\item[$^{21}$] Universidade Federal do Paran\'a, Setor Palotina, Brazil
\item[$^{22}$] Universidade Federal do Rio de Janeiro (UFRJ), Instituto de F\'\i{}sica, Brazil
\item[$^{23}$] Universidade Federal Fluminense, Brazil
\item[$^{24}$] Universidad Industrial de Santander, Colombia
\item[$^{25}$] Institute of Physics (FZU) of the Academy of Sciences of the Czech Republic, Czech Republic
\item[$^{26}$] Palacky University, RCPTM, Czech Republic
\item[$^{27}$] University Prague, Institute of Particle and Nuclear Physics, Czech Republic
\item[$^{28}$] Institut de Physique Nucl\'eaire d'Orsay (IPNO), Universit\'e Paris-Sud, Univ.\ Paris/Saclay, CNRS-IN2P3, France, France
\item[$^{29}$] Laboratoire de Physique Nucl\'eaire et de Hautes Energies (LPNHE), Universit\'es Paris 6 et Paris 7, CNRS-IN2P3, France
\item[$^{30}$] Laboratoire de Physique Subatomique et de Cosmologie (LPSC), Universit\'e Grenoble-Alpes, CNRS/IN2P3, France
\item[$^{31}$] Bergische Universit\"at Wuppertal, Department of Physics, Germany
\item[$^{32}$] Karlsruhe Institute of Technology, Institut f\"ur Experimentelle Kernphysik (IEKP), Germany
\item[$^{33}$] Karlsruhe Institute of Technology, Institut f\"ur Kernphysik (IKP), Germany
\item[$^{34}$] Karlsruhe Institute of Technology, Institut f\"ur Prozessdatenverarbeitung und Elektronik (IPE), Germany
\item[$^{35}$] RWTH Aachen University, III.\ Physikalisches Institut A, Germany
\item[$^{36}$] Universit\"at Hamburg, II.\ Institut f\"ur Theoretische Physik, Germany
\item[$^{37}$] Universit\"at Siegen, Fachbereich 7 Physik -- Experimentelle Teilchenphysik, Germany
\item[$^{38}$] Gran Sasso Science Institute (INFN), L'Aquila, Italy
\item[$^{39}$] INAF -- Istituto di Astrofisica Spaziale e Fisica Cosmica di Palermo, Italy
\item[$^{40}$] INFN Laboratori Nazionali del Gran Sasso, Italy
\item[$^{41}$] INFN, Gruppo Collegato dell'Aquila, Italy
\item[$^{42}$] INFN, Sezione di Catania, Italy
\item[$^{43}$] INFN, Sezione di Lecce, Italy
\item[$^{44}$] INFN, Sezione di Milano, Italy
\item[$^{45}$] INFN, Sezione di Napoli, Italy
\item[$^{46}$] INFN, Sezione di Roma ``Tor Vergata``, Italy
\item[$^{47}$] INFN, Sezione di Torino, Italy
\item[$^{48}$] Osservatorio Astrofisico di Torino (INAF), Torino, Italy
\item[$^{49}$] Universit\`a del Salento, Dipartimento di Ingegneria, Italy
\item[$^{50}$] Universit\`a del Salento, Dipartimento di Matematica e Fisica ``E.\ De Giorgi'', Italy
\item[$^{51}$] Universit\`a dell'Aquila, Dipartimento di Scienze Fisiche e Chimiche, Italy
\item[$^{52}$] Universit\`a di Catania, Dipartimento di Fisica e Astronomia, Italy
\item[$^{53}$] Universit\`a di Milano, Dipartimento di Fisica, Italy
\item[$^{54}$] Universit\`a di Napoli ``Federico II``, Dipartimento di Fisica ``Ettore Pancini``, Italy
\item[$^{55}$] Universit\`a di Roma ``Tor Vergata'', Dipartimento di Fisica, Italy
\item[$^{56}$] Universit\`a Torino, Dipartimento di Fisica, Italy
\item[$^{57}$] Benem\'erita Universidad Aut\'onoma de Puebla (BUAP), M\'exico
\item[$^{58}$] Centro de Investigaci\'on y de Estudios Avanzados del IPN (CINVESTAV), M\'exico
\item[$^{59}$] Unidad Profesional Interdisciplinaria en Ingenier\'\i{}a y Tecnolog\'\i{}as Avanzadas del Instituto Polit\'ecnico Nacional (UPIITA-IPN), M\'exico
\item[$^{60}$] Universidad Aut\'onoma de Chiapas, M\'exico
\item[$^{61}$] Universidad Michoacana de San Nicol\'as de Hidalgo, M\'exico
\item[$^{62}$] Universidad Nacional Aut\'onoma de M\'exico, M\'exico
\item[$^{63}$] Institute for Mathematics, Astrophysics and Particle Physics (IMAPP), Radboud Universiteit, Nijmegen, Netherlands
\item[$^{64}$] KVI -- Center for Advanced Radiation Technology, University of Groningen, Netherlands
\item[$^{65}$] Nationaal Instituut voor Kernfysica en Hoge Energie Fysica (NIKHEF), Netherlands
\item[$^{66}$] Stichting Astronomisch Onderzoek in Nederland (ASTRON), Dwingeloo, Netherlands
\item[$^{67}$] Institute of Nuclear Physics PAN, Poland
\item[$^{68}$] University of \L{}\'od\'z, Faculty of Astrophysics, Poland
\item[$^{69}$] University of \L{}\'od\'z, Faculty of High-Energy Astrophysics, Poland
\item[$^{70}$] Laborat\'orio de Instrumenta\c{c}\~ao e F\'\i{}sica Experimental de Part\'\i{}culas -- LIP and Instituto Superior T\'ecnico -- IST, Universidade de Lisboa -- UL, Portugal
\item[$^{71}$] ``Horia Hulubei'' National Institute for Physics and Nuclear Engineering, Romania
\item[$^{72}$] Institute of Space Science, Romania
\item[$^{73}$] University of Bucharest, Physics Department, Romania
\item[$^{74}$] University Politehnica of Bucharest, Romania
\item[$^{75}$] Experimental Particle Physics Department, J.\ Stefan Institute, Slovenia
\item[$^{76}$] Laboratory for Astroparticle Physics, University of Nova Gorica, Slovenia
\item[$^{77}$] Universidad Complutense de Madrid, Spain
\item[$^{78}$] Universidad de Granada and C.A.F.P.E., Spain
\item[$^{79}$] Universidad de Santiago de Compostela, Spain
\item[$^{80}$] Case Western Reserve University, USA
\item[$^{81}$] Colorado School of Mines, USA
\item[$^{82}$] Colorado State University, USA
\item[$^{83}$] Department of Physics and Astronomy, Lehman College, City University of New York, USA
\item[$^{84}$] Fermi National Accelerator Laboratory, USA
\item[$^{85}$] Louisiana State University, USA
\item[$^{86}$] Michigan Technological University, USA
\item[$^{87}$] New York University, USA
\item[$^{88}$] Northeastern University, USA
\item[$^{89}$] Ohio State University, USA
\item[$^{90}$] Pennsylvania State University, USA
\item[$^{91}$] University of Chicago, USA
\item[$^{92}$] University of Hawaii, USA
\item[$^{93}$] University of Nebraska, USA
\item[$^{94}$] University of New Mexico, USA
\item[] -----
\item[$^{a}$] School of Physics and Astronomy, University of Leeds, Leeds, United Kingdom
\item[$^{b}$] Max-Planck-Institut f\"ur Radioastronomie, Bonn, Germany
\item[$^{c}$] also at Vrije Universiteit Brussels, Brussels, Belgium
\item[$^{d}$] now at Deutsches Elektronen-Synchrotron (DESY), Zeuthen, Germany
\item[$^{e}$] SUBATECH, \'Ecole des Mines de Nantes, CNRS-IN2P3, Universit\'e de Nantes
\end{description}

}
\end{center}

\vspace*{0.3cm}


\begin{abstract}
\noindent

We report a multi-resolution search for anisotropies in the arrival directions of cosmic rays detected at the Pierre Auger Observatory with local zenith angles up to $80^\circ$ and energies in excess of 4~EeV ($4 \times 10^{18}$~eV). This search is conducted by measuring the angular power spectrum and performing a needlet wavelet analysis in two independent energy ranges.  Both analyses are complementary since the angular power spectrum achieves a better performance in identifying large-scale patterns while the needlet wavelet analysis,  considering the parameters used in this work,  presents a higher efficiency in detecting smaller-scale anisotropies, potentially providing directional information on any observed anisotropies.   No deviation from isotropy is observed on any angular scale in the energy range between 4 and 8~EeV. Above 8~EeV, an indication for a dipole moment is captured; while no other deviation from isotropy is observed for moments beyond the dipole one.  The corresponding $p$-values obtained after accounting for searches blindly performed at several angular scales, are $1.3 \times 10^{-5}$ in the case of the angular power spectrum,  and $2.5 \times 10^{-3}$ in the case of the needlet analysis.  While these results are consistent with previous reports making use of the same data set, they provide extensions of the previous works through the thorough scans of the angular scales. 
 
\end{abstract}

\section{Introduction}
\label{sec:intro}

\quad The study of anisotropies in the arrival directions of Ultra-High Energy Cosmic Rays (UHECRs) as a function of energy is a very important element in elucidating their origin. Identifying such anisotropies is not a simple task since they are weakened by the deflection of cosmic rays in Galactic and extragalactic magnetic fields. The study is further complicated by the fact that the strength of the deflection  depends on the cosmic-ray mass composition. 

Such anisotropies may occur over a wide range of angular scales, and dedicated analysis tools are required to capture and summarise the main features of the angular distributions. In the TeV-PeV energy range for instance, Galactic Cosmic Rays (CRs) show complex patterns revealed thanks to the large statistics collected in the last decade by dedicated experiments in both hemispheres. At large scales, anisotropy contrasts at the $10^{-4}-10^{-3}$ level are now well established~\cite{Tibet2005,SK2007,Milagro2008,EASTOP2009,IceCube2010,IceTop2012}\footnote{For a recent review, see \textit{e.g.}, ref. \cite{DiSciascio2014}.}. The dipole moment is generally considered with special interest to probe the particle density gradient shaped by the diffusion of CRs in interstellar magnetic fields on scales of the scattering diffusion length. However, additional effects may enter into play to shape the dipole parameters actually observed, such as the anisotropic diffusion induced by the local ordered magnetic field whose strength is such that circular motions predominate over random scattering. In this case, the observed dipole results from the projection of the density gradient of CRs onto the direction of the ordered magnetic field~\cite{Mertsch2015,Ahlers2016}. Besides, smaller but significant anisotropy contrasts at the $10^{-5}-10^{-4}$ level have also been captured at intermediate and smaller scales~\cite{Milagro2008,IceCube2011,ARGO2013,HAWC2014,IceCube2016}. These complex patterns whose positions appear randomly distributed may be the consequence of the particular structure of the turbulent magnetic field within the last sphere of diffusion encountered by CRs. With a density gradient, this turbulence is indeed expected to connect regions of higher density to regions of lower density as seen from Earth, and \textit{vice versa}~\cite{Giacinti2012,Ahlers2014}. The angular power spectrum, as measured by the IceCube Observatory~\cite{IceCube2011,IceCube2016}, is then a natural tool to capture the main output of the complex stochastic process producing the arrival direction distribution of CRs, and may help in the future to model and constrain the local turbulent magnetic field~\cite{LopezBarquero2016}. Note also that the power spectrum at small scales measured at TeV energies may also highlight relevant signatures to probe and study the electric field induced from the motion of the heliosphere relative to the plasma rest frame where the electromagnetic field can be considered as purely magnetic due to the high conductivity of the medium~\cite{Drury2013}.

Overall, the measured angular power spectrum is thus, nowadays, a valuable analysis to probe and understand the propagation of TeV-PeV CRs in the local  environment, in terms of both the regular and turbulent magnetic fields. This is essentially because invoking a random nature for the arrival direction distributions of CRs allows for a stochastic modeling of non-well known or unknown source positions and propagation mediums; in opposition to CMB studies where the power spectrum results from primordial and fundamental fluctuations.

At higher energies, the expected increase of anisotropy does not compensate, yet, the decrease in the collected statistics with increasing energy, so that only a few indications of dipolar patterns have been reported. Among these indications are $i)$ a consistency of the phase measurements in ordered energy intervals showing a transition at EeV energies that may be indicative of genuine signals~\cite{Auger2011a,AlSamarai2015}, $ii)$ an amplitude of the first harmonic in right ascension standing out from the noise with a $p$-value of $6.4\times10^{-5}$ above 8~EeV in the regions of the sky covered by the Auger Observatory, translating into a dipolar amplitude of $(7.3\pm1.5)\%$ under the assumption that the only significant contribution to the anisotropy is a dipolar pattern~\cite{Auger2015a}, and $iii)$ a dipolar signal with an amplitude $d = \left(6.5 \pm 1.9 \right)\%$ standing out from isotropic expectations with a $p$-value of $5\times10^{-3}$ above 10~EeV obtained with full-sky coverage from a joint analysis of data collected at the Auger Observatory and the Telescope Array~\cite{AugerTA2014,AugerTAICRC2015}. 

Dipolar fluxes are expected in many scenarios of UHECR origin~\cite{Harari2004,Serpico2006,Harari2010,Giller1980,Berezinsky1990,Tinyakov2015}. Quadrupolar type anisotropies are also expected in the case of an excess along a plane such as the Galactic or super-galactic planes, which are plausible source regions at high energies. Although the actual sources of extragalactic UHECRs are still to be identified, their distribution in the sky is expected to follow, to some extent, the large-scale structure of the matter in the Universe. The angular distribution of UHECRs is then expected to be closely connected to the sources and the propagation mode, and to be influenced by the contribution of nearby sources. The Milky Way could thus be, to first order, embedded into a density gradient of UHECRs that should lead to at least a dipole moment. Even for a pure dipole gradient at the entrance of the Galaxy, magnetic deflections are expected to give rise to higher-order multipoles, although with small amplitude~\cite{Harari2010}. This, in combination with possible contributions from random configurations of point sources to the cosmic-ray flux that could also be affected by random configurations of magnetic deflections, shows the importance of a multi-resolution analysis of the arrival directions of UHECRs~\cite{Isola2002,Sigl2003,Sigl2004,Armengaud2005}. In this work therefore, we extend previous anisotropy searches to higher multipoles by measuring the UHECR angular power spectrum and  performing an analysis, based on the needlet wavelet~\cite{Marinucci2008} for energies in excess of the saturation energy of detection efficiency when considering events with zenith angles up to $80^\circ$, namely above 4~EeV. The benefit of using two methods in this paper is that both of these, widely used in astrophysical analyses, are complementary. While the angular power spectrum achieves a better performance in identifying large-scale patterns, the needlet-wavelet analysis method, considering the parameters used in this work, presents a higher efficiency in detecting smaller-scale anisotropies, potentially providing directional information on any observed anisotropies. We note that the needlet-wavelet analysis method could achieve a similar or potentially higher dipole efficiency by adjusting the parameters. However this would be at the cost of sensitivity to smaller scales which would go against the intention of this work.

The paper is organised as follows:  the angular power spectrum and the needlet analysis are presented in Section \ref{sec:methods} with their global estimators to assess the anisotropies in different angular scales, taking into account the incomplete and non-uniform sky coverage. The data set recorded by the Pierre Auger Observatory and used in this work is described in Section \ref{sec:Dataset}. The results of both analyses are presented in Section \ref{sec:global_search}, while the conclusions of this work are discussed in Section \ref{sec:conclusion}.

\section{Anisotropy searches in different angular scales}
\label{sec:methods}

\quad To uncover possible anisotropies in the UHECR flux, we perform two complementary multi-resolution analyses: the widely-used  angular power spectrum analysis, and a wavelet analysis previously used to search for patterns in the cosmic microwave background \cite{Marinucci2008, Feeney2011, Pietrobon2008}. Both analyses as well as their global estimators are briefly presented in the next two subsections.

\subsection{Angular power spectrum} \label{APS}

\quad As emphasised in the introduction, the flux of cosmic rays in the energy range analysed in this paper is known to be isotropic within the sensitivity of the previous and current observatories located in both hemispheres and dedicated to study EeV and trans-EeV cosmic rays, exception made, if confirmed by future data, of a relatively small dipole with an amplitude around $6.5$\% above $\simeq 8-10~$EeV. The flux of cosmic rays can thus be considered, to first order, as essentially isotropic over the entire sphere with eventual small anisotropies. In this regard, it is convenient to decompose the angular distribution of cosmic rays observed by an experiment in some direction $\mathbf{n}$, $\Phi(\mathbf{n})$, by separating the dominant monopole contribution from the anisotropic one, $\Delta(\mathbf{n})$, as
\begin{equation}
\Phi(\mathbf{n})=\frac{N}{4\pi f_1}W(\mathbf{n})\left[ 1+\Delta(\mathbf{n}) \right],
\end{equation}
where $W(\mathbf{n})$ is the relative coverage of the observatory varying from 0 to 1, $f_1=\int \dif\mathbf{n}~W(\mathbf{n})/4\pi$ the fraction of the sky effectively covered by the observatory, and $N$  the total number of observed cosmic rays. In this way, the multipolar expansion of $\Delta(\mathbf{n})$ into the spherical harmonics basis $Y_{\ell m}(\mathbf{n})$ reads
\begin{equation}
\Delta(\mathbf{n}) = \sum_{\ell>0}\sum_{m=-\ell}^{\ell} a_{\ell m} Y_{\ell m}(\mathbf{n}),
\end{equation}
where the $a_{\ell m}$ coefficients encode any anisotropy fingerprint. 

The partial-sky coverage of the Auger Observatory encoded in the $W(\mathbf{n})$ function prevents the multipolar moments $a_{\ell m}$ to be recovered in a direct way through the customary recipe making use of the orthogonality of the spherical harmonics basis~\cite{Sommers2001}. Indirect procedures have to be used, one of them consisting in considering first the `pseudo'-multipolar moments 
\begin{equation}
\label{eqn:convol}
\tilde{a}_{\ell m} = \int \dif\mathbf{n}~W(\mathbf{n})\Delta(\mathbf{n})Y_{\ell m}(\mathbf{n}),
\end{equation}
and then the system of linear relationships relating these pseudo moments to the real ones. Assuming a bound $\ell_{\mathrm{max}}$ beyond which $a_{\ell m}=0$, these relations can be inverted allowing the recovering of the moments $a_{\ell m}$. However, the obtained resolution on each moment does not behave as $\sqrt{K/N}$ (with $K$ a numerical factor depending on $W$, $K=4\pi$ for $W(\mathbf{n})=1$ for instance) as expected from naive statistical arguments, but increases exponentially with the particular bound $\ell_{\mathrm{max}}$ assumed to truncate the multipolar expansion\footnote{Formally, Equation~(\ref{eqn:convol}) implies that the multipolar moments $a_{\ell m}$ are related to the pseudo-multipolar moments $\tilde{a}_{\ell m}$ through a convolution kernel whose coefficients depend only on the $W(\mathbf{n})$ function:
\begin{equation}
K_{\ell m\ell'm'} = \int \dif\mathbf{n}~W(\mathbf{n})Y_{\ell m}(\mathbf{n})Y_{\ell' m'}(\mathbf{n}).
\end{equation}
The resolution on each recovered moment is then $\sqrt{4\pi f_1[K^{-1}]_{\ell m\ell'm'}/N}$, the inverse matrix $K^{-1}$ depending on the bound $\ell_{\mathrm{max}}$.
}~\cite{Billoir2008}.  

Hence, given the current exposure of the Observatory to cosmic rays above 4~EeV, the estimation of the individual $a_{\ell m}$ coefficients cannot be carried out with relevant resolution as soon as $\ell_{\mathrm{max}}>2$. However, based on analysis techniques previously developed in the CMB community~\cite{Hivon2002}, it is possible, under some restrictions detailed now and further discussed below, to reconstruct the angular power spectrum coefficients ${C}_{\ell}=\sum^{\ell}_{m=-\ell}|a_{\ell m}|^{2}/(2\ell +1)$ within a statistical resolution independent of the bound $\ell_{\mathrm{max}}$~\cite{Deligny2004}. The starting point is to consider that the observed distribution of arrival directions is a particular realisation of an underlying stochastic process which can be assumed Gaussian in a conservative way, and which is thus entirely characterised by its first two moments $\langle\Delta(\mathbf{n})\rangle$ and $\langle\Delta(\mathbf{n})\Delta(\mathbf{n'})\rangle$. In the absence of knowledge of this stochastic process, the simplest non-trivial situation is to consider that the anisotropies cancel in ensemble average and produce a second order moment that does not depend on the position on the sphere but only on the angular separation between $\mathbf{n}$ and $\mathbf{n'}$. In this case, the underlying $a_{\ell m}$ coefficients vanish in average and are not correlated to each other (\textit{i.e.} diagonal covariance: $\langle a_{\ell m}a_{\ell' m'}\rangle=C_\ell\delta_{\ell\ell'}\delta_{mm'}$) so that the ${C}_{\ell}$ coefficients can be viewed as a measure of the variance of the $a_{\ell m}$ coefficients. In this situation, it can then be shown that the pseudo-power spectrum $\tilde{C}_{\ell}=\sum^{\ell}_{m=-\ell}|\tilde{a}_{\ell m}|^{2}/(2\ell +1)$ (which is directly measurable) is related to the real power spectrum through 
\begin{eqnarray}
\label{eqn:Cl}
\tilde{C}_{\ell} = \sum_{\ell'}M_{\ell\ell'}C_{\ell'},
\end{eqnarray}
where the operator $M$ describing the cross-talk induced by the non-uniform exposure between genuine modes is entirely determined by the knowledge of the exposure function. The power spectrum can thus be recovered from the inversion of $M$. Interestingly, for a two-point function of the exposure\footnote{The two-point function of the exposure is the average over position and orientation on the sphere of $W$ at angular separation $\gamma$ defined as 
\begin{equation}
\mathcal{W}(\cos\gamma) = \int\int\frac{\dif\mathbf{n}\dif\mathbf{n'}}{8\pi^2}W(\mathbf{n})W(\mathbf{n'})\delta(\mathbf{n}\cdot\mathbf{n'}-\cos\gamma),
\end{equation}
the normalization factor being chosen such that $\mathcal{W}=1$ for $W=1$.} $\mathcal{W}(\cos\gamma)$ never vanishing as in the case of the Auger Observatory, $M^{-1}$ is shown in the Appendix to take the form 
\begin{equation}
\label{eqn:Minv}
M^{-1}_{\ell\ell'} = \frac{2\ell'+1}{2}\int_{-1}^1\frac{\dif\cos{\gamma}}{\mathcal{W}(\cos\gamma)}~P_\ell(\cos{\gamma})P_{\ell'}(\cos{\gamma}), 
\end{equation}
where the standard notation $P_\ell(\cos{\gamma})$ stands for the Legendre polynomials. Under this form, it is explicit that the inversion operation of Equation~(\ref{eqn:Cl}) is unambiguously defined, independently of the bound $\ell_{\mathrm{max}}$. 

We now discuss in more details the implications of the hypothesised two first moments of the stochastic field modeling the angular distribution of cosmic rays, namely $\langle\Delta(\mathbf{n})\rangle=0$ and $\langle\Delta(\mathbf{n})\Delta(\mathbf{n'})\rangle=\zeta(\mathbf{n}\cdot\mathbf{n'})$ (with $\zeta$ any function). These conditions are often referred to as conditions of stationarity and expressed in the reciprocal space as $\langle a_{\ell m}\rangle=0$ and $\langle a_{\ell m}a_{\ell' m'}\rangle=C_\ell\delta_{\ell\ell'}\delta_{mm'}$. Although the underlying process governing the stochastic distribution of the $a_{\ell m}$ coefficients is unknown, these conditions can describe or approximate numerous benchmark scenarios of interest. One of these benchmark scenarios, widely-used in the literature, relies on drawing sources at random with a granularity in accordance with some density that might depend on the redshift. Whatever the propagation regime, the ensemble average angular distribution is isotropic by construction and the particular anisotropies in each realisation are due to the fluctuations of the positions of the most contributing local sources. Even with some types of structure for the local sources on top of the contribution of the numerous distant sources, the random localisation of the observer and a diffusive propagation regime allow a stationarity description of the process, to first order at least. On the other hand, since the stationarity conditions are obviously not comprehensive of all stochastic processes, there are scenarios preventing the power spectrum to be fairly captured with the method used here. For instance, one can think of an observer and sources randomly distributed within a thick disk on local scales. In the case of a ballistic or quasi-ballistic propagation regime, some moments, predominantly the quadrupolar ones, would not average to zero. Another example of scenario, that would induce significant non-zero correlations between several moments and thus break the condition $\langle a_{\ell m}a_{\ell' m'}\rangle=C_\ell\delta_{\ell\ell'}\delta_{mm'}$, is the one of a unique `hot-spot' at some small or intermediate angular scale on top of an isotropic distribution on the whole sphere. These kinds of scenarios would be better probed and captured by other analysis techniques, such as the wavelet one presented farther.

Overall, given the coverage of the sky at our disposal for this study, making use of the stationarity conditions requires the assumption of the absence of intermediate-scale and small-scale structures in the uncovered region of the sky. Since the sensitivity of previous or current experiments covering the Northern hemisphere is smaller than the one reached in this report (for instance, $\simeq5.5$ times more exposure here than in the recent search performed at the Telescope Array~\cite{Abbasi2016}), the non-detection of such structures in the Northern hemisphere\footnote{The focus here is on energies between 4 and 8~EeV and above 8~EeV, so the discussion is not affected by the hot spot above 57~EeV reported by the Telescope Array Collaboration in~\cite{TA-hotspot} due to the steepness of the energy spectrum.},  as reported in energy ranges covering the ones considered in this study for instance in~\cite{Bird1999, Abbasi2004, Abu-Zayyad, AugerTA2014},  does not prevent the possibility of a hot spot with an amplitude below the current limits. In this case, the 'hidden' hot spot would influence to a larger extent the large-scale moments, and the stationarity conditions would tend to cancel some multipoles, essentially the quadrupolar ones. In this sense, an unbiased estimation of the power spectrum requires full-sky coverage. 


To facilitate the introduction to the angular power spectrum technique, only idealised quantities have been presented up to here. The finite sampling of the angular distribution induces Poisson fluctuations and requires the introduction of the following estimator for the $\tilde{a}_{\ell m}$ coefficients~\cite{Deligny2004}:
\begin{equation}
\tilde{a}_{\ell m,\mathrm{data}} = \int_{4\pi} \dif\mathbf{n}~Y_{\ell m}(\mathbf{n})\frac{\dif N/\dif\mathbf{n}-(N/4\pi f_1)W(\mathbf{n})}{N/4\pi f_1}, 
\end{equation}
where $\dif N/\dif\mathbf{n}$ is the observed distribution of arrival directions. In terms of statistical performances, the Poisson fluctuations induce an irreducible noise on the estimated power spectrum coefficients such that
\begin{equation} 
\label{bias_iso}
C_{\ell,\mathrm{data}} = C_\ell+\frac{4\pi}{N}\frac{f_1^2}{f_2},
\end{equation}
with $f_2=\int\dif\mathbf{n}~W^2(\mathbf{n})/4\pi$; while for isotropic samples the resolution obtained on each recovered power for each mode $\ell$ behaves as
\begin{equation} 
\label{variance_iso}
\sigma(C_{\ell,\mathrm{data}}) = \left(  \frac{4 \pi f_1}{N}\right) \sqrt{\frac{2 }{2\ell+1} M^{-1}_{\ell \ell}}.
\end{equation}
Note that to obtain these uncertainties, stationarity conditions are used explicitly. 

Besides searching for anisotropies at each individual scale, it is also important to define a global anisotropy estimator, which computes the departure of all observed $C_{\ell}$ in relation to those expected from an isotropic distribution and at the same time penalises statistically the search over many angular scales. This estimator, inspired by~\cite{Hulss2007}, is given by
\begin{equation}
D^{2}=\frac{1}{\ell_{\mathrm{max}}}\sum_{\ell =1}^{\ell_{\mathrm{max}}}\left(\frac{C_{\ell , \mathrm{data}}-\langle C_{\ell , \mathrm{iso}}\rangle}{\sigma_{\ell , \mathrm{iso}}}\right)^{2},
\label{eq:estimator}
\end{equation}where $C_{\ell, \mathrm{data}}$, $\langle C_{\ell, \mathrm{iso}} \rangle$ and $\sigma_{\ell, \mathrm{iso}}$ are, respectively,  the $C_{\ell}$  observed in the data, the average and standard deviation from isotropic $C_{\ell}$ expectations, all of them evaluated at a given scale $\ell$. In practice, we obtain the $\langle C_{\ell, \mathrm{iso}} \rangle$ and $\sigma_{\ell, \mathrm{iso}}$ values using simulated isotropic skies with the same number of events and exposure as for the data. The choice of a Confidence Level (C.L.) of 99\% defines a threshold  $D^2_{\mathrm{iso}, 99\%}$ to accept or reject the isotropy hypothesis: distributions whose values of $D^2$ are lower (larger) than this $D^2_{\mathrm{iso}, 99\%}$  are considered isotropic (anisotropic) at the 99$\%$ C.L..


\subsection{Needlet method}
\label{Needlet}

\quad The needlet is a type of spherical wavelet that was introduced to search for patterns in the cosmic microwave background (CMB)~\cite{Marinucci2008}. The needlet may be viewed as a filter which enhances existing global and localised anisotropies of a given size. The needlet is composed of multiple scales (\textit{i.e.} filters), which are sensitive to angular structures of different sizes, ranging from large-scale to small-scale structures. Among the variety of wavelets implementations, the properties of the needlets are particularly well suited for the study pursued here~\cite{Marinucci2008}: they are exactly localised on a finite number of multipoles that can be specified for a particular analysis, they are quasi-exponentially concentrated in pixel space, and they are directly embedded on the sphere and thus do not rely on a tangent plane approximation, similarly to the wavelet implementation used in~\cite{Zimbres2013}.
The principle of the analysis presented below is thus to expand the data set onto spherical harmonics and to convolve the expansion with the different needlet scales so that anisotropies of various sizes can be found. The end result lies in an output sky map in which existing anisotropies are enhanced. 

As in the case of the angular power spectrum approach, the pseudo-spherical harmonic coefficients $\tilde{a}_{\ell m}$ are used. These harmonic coefficients are derived in the same way with the exception that the event map is weighted by the inverse of the Auger coverage (where the coverage is non-zero) before expansion. For the representation of the data on the sphere and for the harmonic expansion the HEALPix package~\cite{Healpix} is used. The coefficients are then convolved with the needlet wavelet, renormalised and transformed back into pixel space. The result of the analysis is a filtered sky map $S_k$ from which a global anisotropy estimator $S$ is derived (cf. Equation~(\ref{eq:estimator})). The process is described in more detail in the following.

\begin{figure}[hbtp]
  \begin{center}
  \includegraphics[width=.9\textwidth]{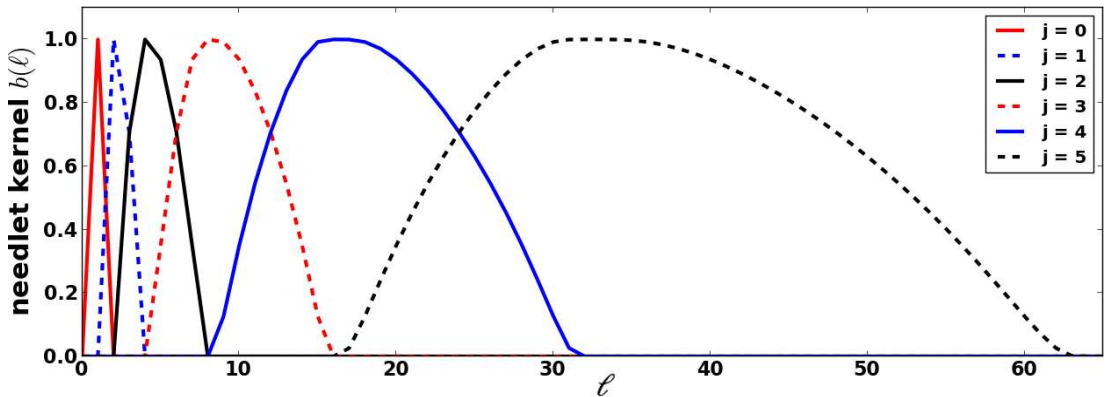}
  \caption[Needlet kernel]{Needlet kernel $b(\ell,(B=2.0)^{-j})$ as a function of the multipole moment $\ell$ at different needlet scales $j$. 
  }
  \label{fig:Needlets1B}
 \end{center}
\end{figure}

Needlets may intuitively be viewed as a convolution of the projection operator $ \sum_{m=-\ell}^\ell \bar{Y}_{{\ell m}}(\theta,\phi)Y_{{\ell m}}(\theta_k,\phi_k)$, where $\theta_k$ and $\phi_k$ denote the coordinates of the centre of the pixel $k$ of the HEALPix scheme (see Equation~(\ref{eqn:PixelPowerForm})),  with a suitably-chosen window function $b(.)$. In our case this is the needlet kernel\footnote{For details on the construction of the needlet kernel see~\cite{Marinucci2008}.} $b(\ell, B^{-j})$ with its different the scales $j$, which is shown in Figure~\ref{fig:Needlets1B} as a function of the multipole moment $\ell$. At a fixed maximum multipole moment, the width of the needlet scales and consequently the number of available $j$-scales is determined by the parameter $B$.  The properties of $b$ ensure a quasi-exponential localization property of the needlet~\cite{Marinucci2008}.

A value of $B=2.0$ is used in the analysis which gives a reasonable range of scales (six in our case, see Figure~\ref{fig:Needlets1B} and Section~\ref{sec:NeedRes}) and is a good trade-off between a tighter localization in harmonic space versus in pixel space. This choice is based on extensive Monte-Carlo sensitivity studies to various kinds of anisotropy  scenarios such as dipole, quadrupole, catalogue-based and point-source scenario\footnote{The same holds true for the parameter choice in Equation~(\ref{eqn:NThreshold}).}. We choose the respective values to give an overall good sensitivity to detect anisotropies across the various scenarios. For example, a lower value of the (threshold) parameter in Equation~(\ref{eqn:NThreshold})  leads to a better sensitivity to large-scale structures but to a lower sensitivity to small-scale, localised structures such as point sources. As can be seen in Figure \ref{fig:Needlets1B}, small $j$-scales are sensitive to large-scale structures and \textit{vice versa}.

The needlet is convolved with the (pseudo) harmonic coefficients for each scale $j$ individually. After the convolution a backwards transformation is performed, resulting in a set of `power' sky maps $\beta_{jk}$. The signal power $\beta$ in each pixel $k$ (in the HEALPix scheme) and scale $j$ is then given by \cite{Pietrobon2006}:
\begin{equation}
  \beta_{jk} = \sqrt\lambda~\sum_\ell b(\ell,B^{-j}) \sum_{m=-\ell}^\ell \tilde{a}_{{\ell m}} Y_{{\ell m}}(\theta_k,\phi_k).
  \label{eqn:PixelPowerForm}
\end{equation}
Here $\theta_k$ and $\phi_k$ denote the coordinates of the centre of the pixel $k$ of the HEALPix scheme and $\sqrt\lambda$ is a normalisation factor given by $\lambda = 1/N_\text{pix}$, where $N_\text{pix}$ is the number of pixels on the sphere.

To combine the different scales and to search for deviations from isotropy in the sky map, each individual scale $j$ is normalised as described in the following:
\begin{itemize}
\item  First, a large set of isotropic sky maps with a given number of observed events is simulated as would be detected by an observatory with a given exposure.
\item  Second, from this set the mean pixel power $\langle\beta_{jk,\mathrm{iso}}\rangle$ and the pixel power fluctuations $\sigma_{jk,\mathrm{iso}}$ of each pixel $k$ in each scale $j$ are determined.
In the case of a uniform sky exposure, the means and the variances would be identical for all pixels. On the other hand, 
 in the case of non-uniform/incomplete exposure, they depend on the declination and/or right ascension.
\item Third, each pixel power value is replaced by the pixel significance (as in \cite{Pietrobon2008})
\begin{equation}
  S_{jk} = \frac{|\beta_{jk}-\langle\beta_{jk,\mathrm{iso}}\rangle|}{\sigma_{jk,\mathrm{iso}}} \times \operatorname{sgn}(\beta_{jk}),
\end{equation} where $\operatorname{sgn}(\beta_{jk})$ is the signum function defined as follows:
\begin{equation*}
\operatorname{sgn}(\beta_{jk})=\begin{cases}
					-1 & \text{ if  $\,\beta_{jk} < 0$},\\
					 0 & \text{ if  $\,\beta_{jk}= 0$},\\
					 1 & \text{ if  $\,\beta_{jk} > 0$}.
					\end{cases}
\end{equation*}

\item  Fourth, to reduce the still remaining background in the individual scales, each pixel which satisfies
\begin{equation}
  |S_{jk}| < 3
  \label{eqn:NThreshold}
\end{equation}
is assigned a value of zero. The result is a normalised, filtered set of sky maps $S_{jk}$. As with our choice of the needlet width $B$, this value is chosen to give good sensitivity to anisotropies  over a wide range of scales, determined using Monte-Carlo simulations.
\item  Fifth, a combined filtered sky map is created by summing up all scales $j$ of interest
\begin{equation}
  S_{k} = \sum_j \ S_{jk}\ .
\end{equation}
Depending on the intended analysis, one may only use one or two scales (\textit{e.g.} $j=0$) or all scales which are sensitive up to a given multipole moment $\ell$ to perform an undirected search, \textit{i.e.} over all angular scales, for anisotropy.
\item  Finally, to determine the anisotropy level of a given filtered sky map, we define the $S$ anisotropy estimator in analogy to the $D^2$  estimator (Equation (\ref{eq:estimator})):
\begin{equation}
  S = \log \left(\frac{\sum_{k=1}^{N_\text{pix}}\ |S_{k}|}{N_{S}}\right), \quad N_{S} > 0,
\end{equation}
where $N_{S}$ is the number of pixels where $S_{k}$ does not vanish.\\
\end{itemize}
The $S$-values of a different set of isotropic test sky maps are combined to create an expectation of the  $S$-value distribution in case of isotropy. 
The method to determine whether a given signal sky map deviates from isotropy at a given C.L. is identical to that for the angular power spectrum. Note that when only one single $j$-scale is analysed, no pixel of a given sky map might pass the threshold introduced in Equation~(\ref{eqn:NThreshold}). In this case, a flat distribution in the $S$-distribution of sky maps not passing the cut is assumed.

\section{Data set}
\label{sec:Dataset}

\quad The Pierre Auger Observatory~\cite{Auger2015b}, located in Malarg\"{u}e, Argentina, is the world's largest cosmic-ray observatory. Its design comprises a surface detector (SD) made up of an array of 1660 water-Cherenkov detectors spread over 3000~km$^2$ overlooked by an air-fluorescence detector (FD) comprising a total of 27  telescopes. The SD samples the particle components of extensive air showers on the ground with a duty cycle of nearly $100\%$. Although the FD has a smaller duty cycle of $\sim$15$\%$,  its calorimetric measurement of the shower energy deposited in the atmosphere is very important for the calibration of the SD energy reconstruction.

 The data set used in this work is the same used in~\cite{Auger2015a} for which two Rayleigh analyses, one in the right ascension and one in the azimuth angle distributions, were performed. It is composed of events detected with the SD of the Pierre Auger Observatory from 2004 January 1 to 2013 December 31 with zenith angle $\theta$ up to 80$^{\circ}$. Vertical and inclined events are defined as those events with zenith angle smaller than $60^{\circ}$ and from $60^{\circ}$ up to $80^{\circ}$, respectively.  The set of events was divided in two energy bins: $4\leq E/\mathrm{EeV}<8$ and $E\geq 8$~EeV. This choice for the size of the energy bins, although arbitrary, follows from previous studies dedicated to large-scale anisotropy searches where a bin size such that $\Delta\log_{10}E=0.3$ was selected for $17.4\leq\log_{10}{(E/{\mathrm{eV})}}\leq18.9$. This choice was aimed at guaranteeing a bin size larger than the energy resolution to avoid to weaken the sensitivity of a search for an energy-dependent anisotropy. For $\log_{10}{(E/{\mathrm{eV})}}\geq18.9$, the single bin choice is dictated by the low statistics. With these energy ranges we have exploited the fact that the surface detector of the Pierre Auger Observatory is fully efficient for vertical events with an energy above 3~EeV \cite{Auger2010a} and for horizontal events with an energy above 4~EeV \cite{Auger2015c}, making the determination of the exposure straightforward as systematic uncertainties on trigger and other effects become negligible. The corresponding total exposures are 37,142 km$^2$~sr~yr for the vertical events and 10,887 km$^2$~sr~yr for the inclined events. The total number of events used in this work for both energy ranges are summarised in Table~\ref{table_data_Roger}. 
 
\begin{table}[!htp!]
\begin{center}
\begin{tabular}{c|c|c}
Energy range (EeV) & $\theta < 60^{\circ}$ & $60^{\circ} \leq \theta \leq 80^{\circ}$ \\ \hline \hline
4 - 8 & 39,049 & 11,368 \\ \hline
$>$ 8 & 15,418 & 4,379 \\ 
\end{tabular}
\caption{Number of events for vertical and inclined data sets for different energy ranges used in this work.}
\label{table_data_Roger}
\end{center}
\end{table}

For vertical events, a quality cut requires that all six water-Cherenkov detectors\footnote{A hexagon of water-Cherenkov detectors defines an elemental cell whose geometric aperture, especially important for the exposure calculation, is $a_{\mathrm{cell}}(\theta) = 1.95 \cos \theta $ km$^2$ under incidence at zenith angle $\theta$  \cite{Auger2010a}.} surrounding the station with the largest signal were operational at the time the event was recorded. On the other hand, for inclined events we require that six stations around the station closest to the core position were active at the time of detection. In all cases, the arrival directions of cosmic rays are determined from the relative arrival times of the shower front in the triggered stations.

The primary energy estimator for vertical and inclined events is reconstructed based on simultaneous  measurements from water-Cherenkov detectors and the fluorescence detector forming hybrid events.  For vertical events, the signal value at the ground at a distance of 1000~m from the shower axis, $S(1000)$ is set up as a reference. As $S(1000)$ decreases with zenith angle due to the attenuation of the shower particles and geometrical effects, the shape of the attenuation curve from the data is employed to convert $S(1000)$ to $S_{38}$, that may be regarded as the signal a particular shower with size $S(1000)$ would have produced had it arrived at a zenith angle of $38^{\circ}$. Finally, $S_{38}$ is related to the almost-calorimetric measurements from the FD~\cite{Auger2015b}. The energy reconstruction of an inclined shower is based on the universal shape of the muon distribution and the scaling between muon number density and shower energy~\cite{Auger2014a}. The measured signals are fitted to the expected muon density at the ground via: $\rho_{\mu}(\mathbf{n})=N_{19}\times \rho_{\mu,19}(\mathbf{n})$, where $\mathbf{n}$ is the arrival direction of the primary particle, $N_{19}$ is the shower size and $\rho_{\mu,19}(\mathbf{n})$ is a reference muon distribution, conventionally chosen to be the average muon density for primary protons with energy $10$~EeV simulated with the QGSJetII-03 hadronic interaction package. Using $N_{19}$ as the energy  estimator, the energies of the inclined events are calibrated using a reconstruction similar to the vertical one~\cite{Auger2015b}.

To avoid spurious effects due to array growth as well as the dead periods of each detector, we correct the exposures by considering the variations in the number of active elemental cells. The small tilt of the array of about $0.2^{\circ}$ towards a direction $30^{\circ}$ from the East to the South $(\phi_{\mathrm{tilt}}=-30^{\circ})$ modulates the effective elemental cell area. To correct for these two effects, each event must be weighted by a factor given by
\begin{equation}
\left[\Delta N_{\mathrm{cell}}(\alpha_{0})\right]^{-1}\times \left[1+0.003\text{ tan}\theta \text{ cos}(\phi - \phi_{\mathrm{tilt}})\right]^{-1},
\label{corrections}
\end{equation}
where $\Delta N_{\mathrm{cell}}(\alpha_{0})$ is the relative variation of the total number of active cells as a function of the local sidereal time\footnote{For practical reasons, $\alpha_{0}$ is chosen so that it is always equal to the right ascension of the zenith at the center of the array.}, $\alpha_{0}$, while $\theta$ and $\phi$ are the local arrival direction coordinates of the primary particle \cite{Auger2012a, Auger2013a}. Moreover, the data set for the vertical events has the energy estimator corrected for weather and geomagnetic effects \cite{Auger2012a, Auger2013a} while the energy of the horizontal events has not been corrected for these effects since the geomagnetic field is already accounted for in the shower reconstruction and the weather conditions do not influence the horizontal events as they are dominated by muons at ground level \cite{Auger2014a}. 

  Whether systematic effects affecting anisotropy searches at large scales are under control can be checked by performing a first harmonic analysis for fictitious right ascensions calculated by contracting/dilating the time of the events in such a way that a solar day lasts about four minutes less or longer. The corresponding time scales are the solar and anti-sidereal ones. This is because weather and array size variations have the largest effect in producing spurious modulations at the solar time scale, while a seasonal variation of the solar distortions of the event rate could produce a spurious modulation at the anti-sidereal time scale~\cite{Farley}. Analyses presented in~\cite{Auger2015a} for the two time scales, with the same data set used in the present work, show that no spurious effects appear for any of the energy bins.

\section{Results}
\label{sec:global_search}

\quad The results achieved by measuring the angular power spectrum and by performing a needlet analysis on the Auger data set are described in the next two subsections.   Both analyses can be performed up to an $\ell_{\mathrm{max}}$ corresponding to the  angular resolution of the experiment which is of the order of $\sim 1.5^{\circ}$ for events with $3<E/\mathrm{EeV}<10$ and goes down to $\sim 1^{\circ}$ for events $E>10~$EeV\cite{Bonifazi2009}. This corresponds roughly to an $l_\text{max}$ of 128 and a corresponding Healpix parameter $N_\text{side}$ of 64 resulting in 49,152 pixels on the sphere\footnote{The Healpix $N_\text{side}$ parameter determines the number of pixels in the sphere as N$_\text{pixel} = 12 \times N_\text{side}^2$.}. However, the expected deflection of even the highest energetic protons is around $\sim 3^o$ \cite{Auger-TA-IC} through the galactic magnetic field (GMF) alone, according to the two most recent GMF models \cite{ Pshirkov-Tinyakov, JF12}. Considering this minimum expected smearing of UHECRs and additionally aiming to increase the per-pixel statistics as well as to reduce the computing time, we choose the next lower $N_\text{side}$ resolution of 32. Thus the harmonic expansion is performed up to $\ell_\mathrm{max}=64$ ($\sim 2.8^o$) which marks the end of sensitivity of the needlet scales $j=0-5$ and is also recommended for the chosen resolution~\cite{HealpixDoc}.

\subsection{Angular power spectrum}

\begin{figure}[htp!]
\center
\subfloat{\includegraphics[width=7.7cm,height=5.4cm]{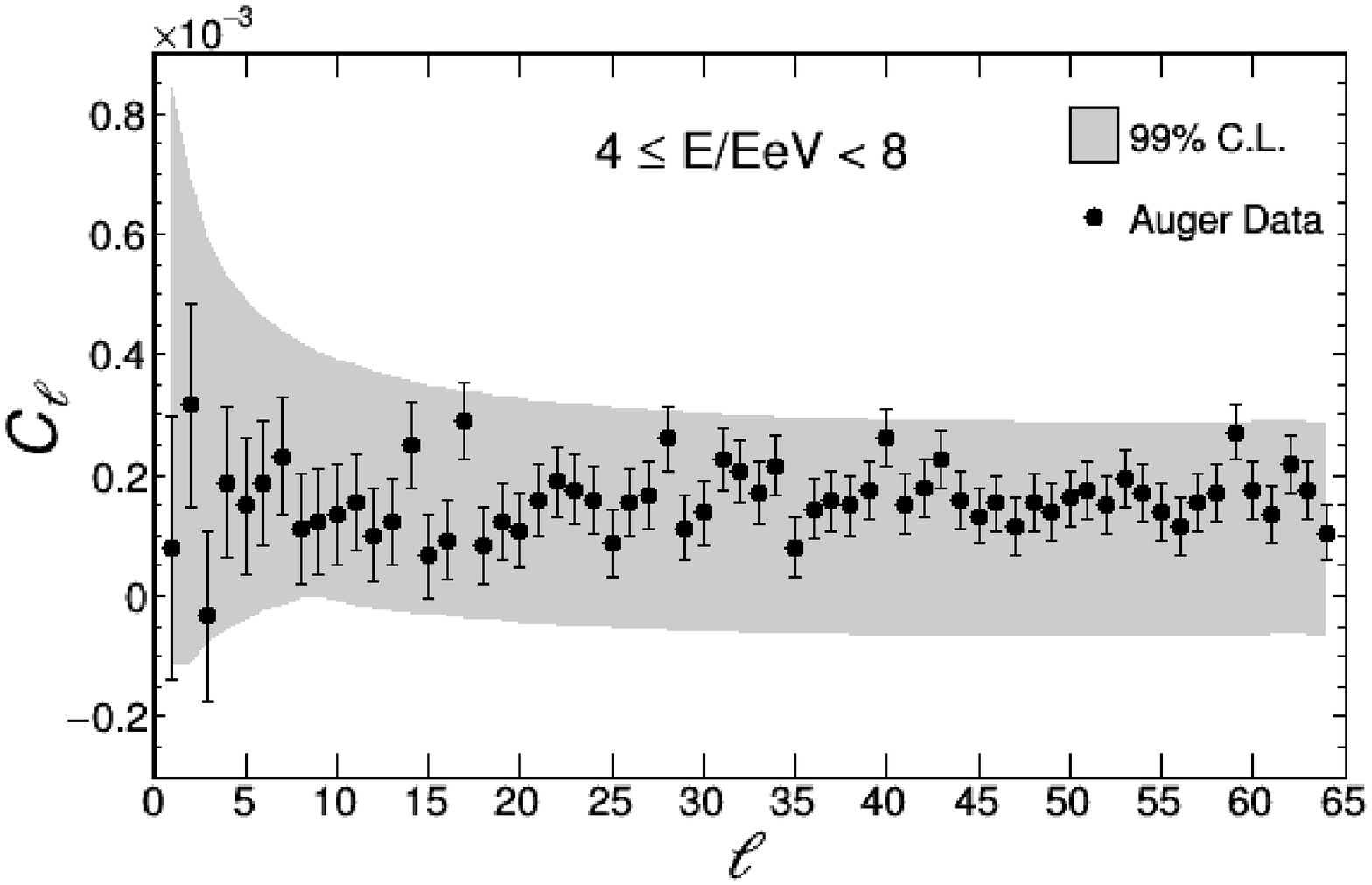}}
\quad
\subfloat{{\includegraphics[width=7.7cm,height=5.5cm]{{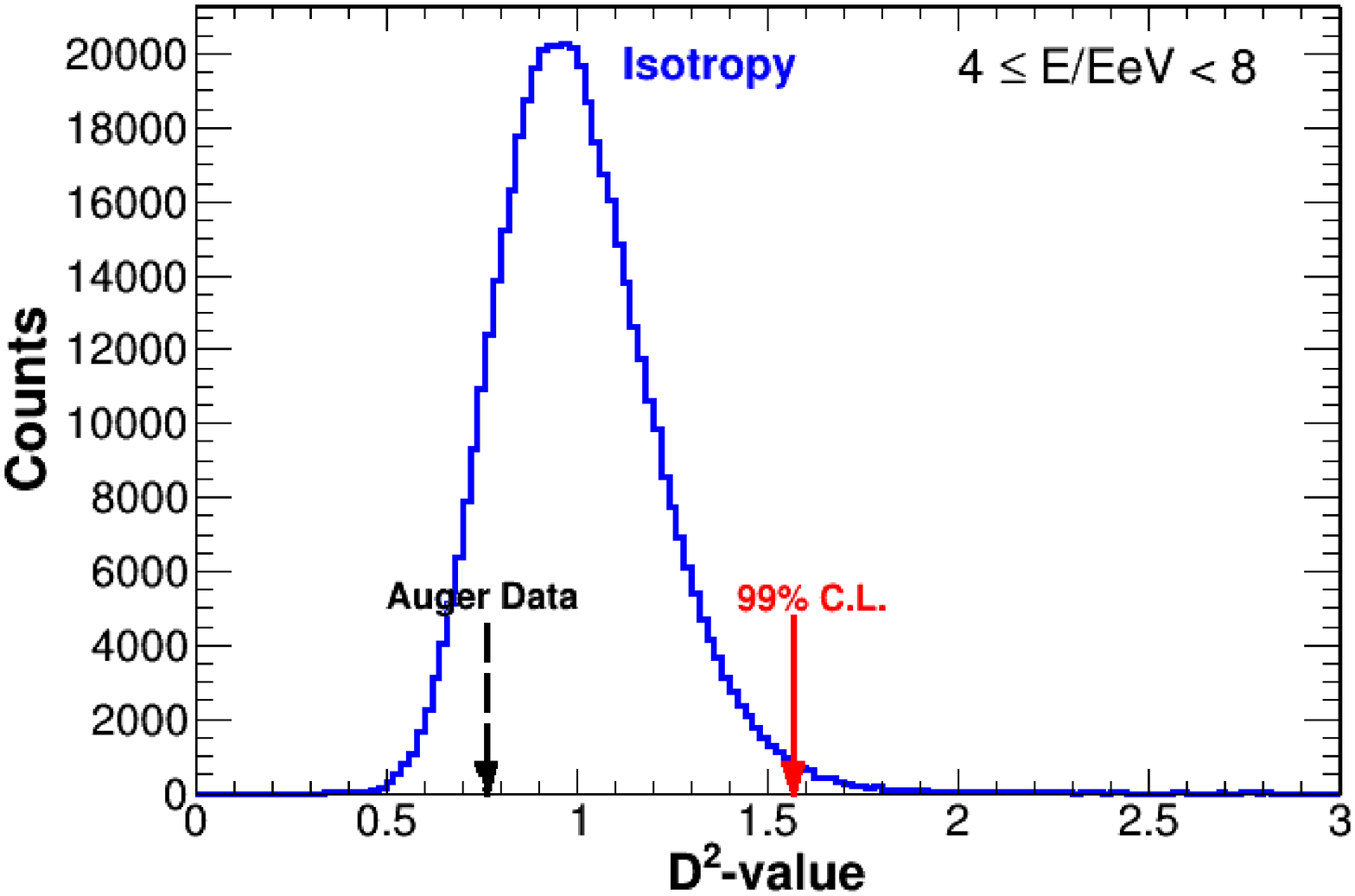}}}}
\caption{Angular power spectrum for $4\leq E/\mathrm{EeV}<8$. On the left there is no visible departure from the isotropic expectation. On the right the $D^{2}$-value distribution from 500,000 isotropic sky maps is shown. The red arrow represents the threshold to accept/reject the isotropy hypothesis with 99\% C.L.. The $D^{2}$-value from data, represented by the black (dashed) arrow, is smaller than that threshold supporting the isotropy hypothesis.}
\label{fig:4-8}
\end{figure}

\quad In the following we evaluate the angular power spectrum in two energy ranges: $4\leq E/\mathrm{EeV}<8$ and $E\geq 8~$EeV. Shown in the left panel of Figure~\ref{fig:4-8} is the angular power spectrum obtained for the first energy bin. The points are observed to fluctuate randomly around the mean noise level expected from Equation~(\ref{bias_iso}), $\langle C_{\ell,\mathrm{iso}}\rangle=1.7\times 10^{-4}$, so that no deviation from isotropy is detected at any scale. The uncertainties in the angular power spectrum are evaluated considering the $C_{\ell}$ variances arising from isotropic distributions given by Equation~(\ref{variance_iso}), which are strictly valid under stationarity conditions. The colored band region, obtained from Monte-Carlo simulations, stands for the 99\% C.L. lower and upper bounds that would result from fluctuations of an isotropic distribution. Note that the negative values are a residual artefact of the incomplete coverage of the sky resulting from negative elements of the $M^{-1}$ matrix coupled to particular fluctuations of the pseudo-power spectrum.

In the right-hand panel, the $D^{2}$-value distribution from 500,000 isotropic simulations is shown. The black arrow shows the $D^{2}$-value of the Auger data set and the red arrow represents the 99\% C.L. under the hypothesis of isotropy. Of all isotropic sky maps about 90\% have either the same or a higher significance (lower $p$-value) and hence, as anticipated from the left-hand panel, no deviation from isotropy is detected.

\begin{figure}[h!]
\center
\subfloat{\includegraphics[width=7.7cm,height=5cm]{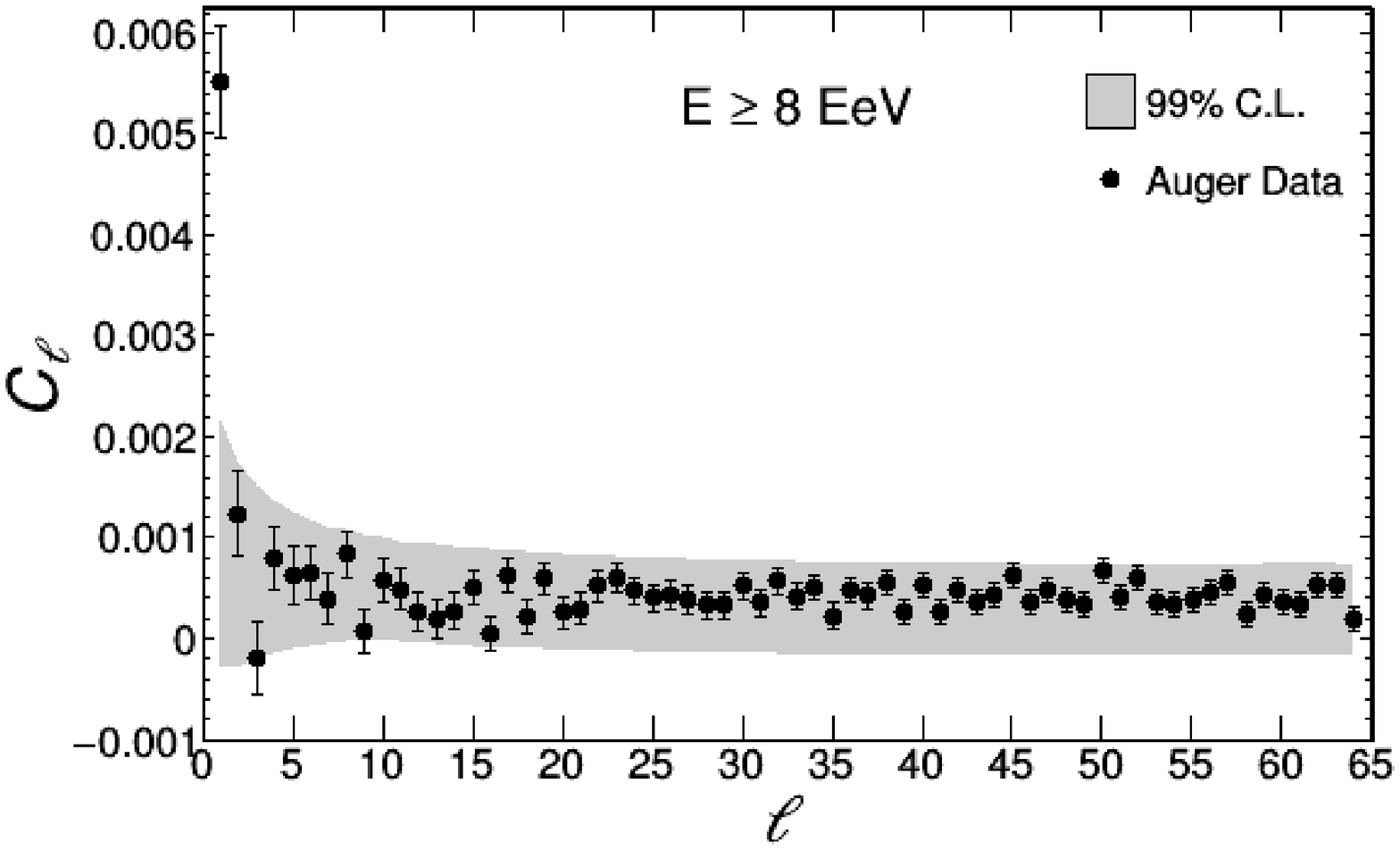}}
\quad
\subfloat{{\includegraphics[width=7.7cm,height=5cm]{{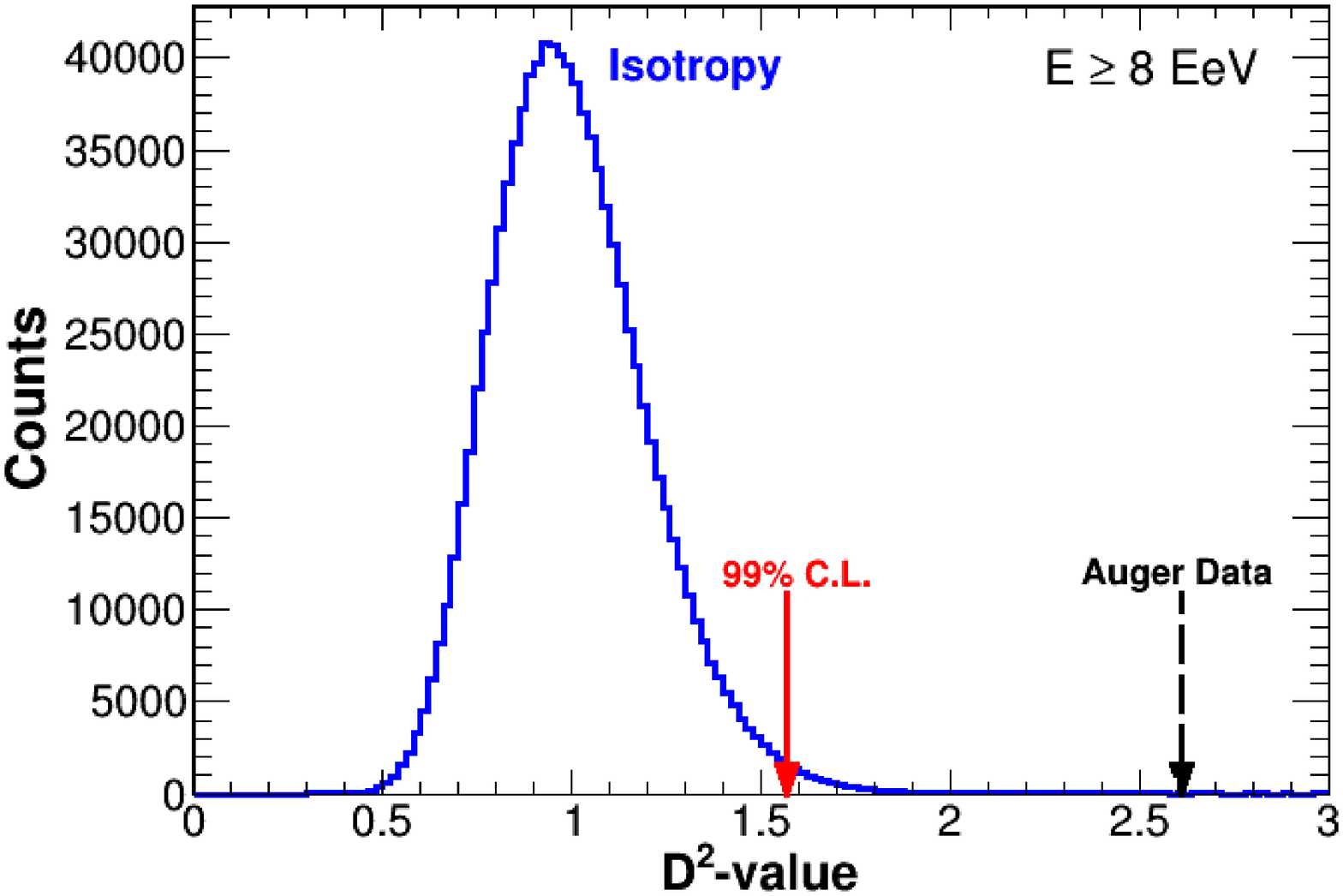}}}}\caption{Angular power spectrum for  $E\ge8$~EeV.  On the left a clear indication for a departure from isotropy is captured in the dipole scale.  On the right the $D^{2}$-value distribution from 1,000,000 isotropic sky maps is shown. The $D^{2}$-value from data, represented by the black (dashed) arrow, is larger than the threshold of isotropy presenting an indication of anisotropy with $>99$\% C.L.. }
\label{fig:8EeV}
\end{figure}

The results for the $E\geq 8$~EeV energy bin are presented in Figure~\ref{fig:8EeV}. The angular power spectrum is shown in the left panel. Under stationarity conditions, no deviation is captured for any multipole moment besides $\ell=1$. Hence, and still under stationarity conditions, the estimated dipole amplitude  $d_{C_1}$ from the $C_1$ value turns out to be
\begin{equation} \label{ampc1}
d_{C_1} =\sqrt{\frac{9\, C_1}{4 \pi}} = (6.0 \pm 1.5) \%,
\end{equation}
where the uncertainty in the dipole amplitude was obtained from simulations, setting the input value of $C_1 = 0.0050$ and $C_\ell=0$ for all $\ell>1$, resulting in the specific value  $C_1=0.0050 \pm 0.0025$. Although this does not constitute a validation of the stationarity assumption, it is to be noted that this amplitude is consistent within uncertainties with the estimates reported in \cite{Auger2015a} under the assumption of a dipolar or dipolar/quadrupolar flux and reported in \cite{AugerTAICRC2015}.  As previously mentioned, the analysis reported in \cite{Auger2015a} is based on the two Rayleigh analyses of the same data set used in this work whereas the one reported in \cite{AugerTAICRC2015} is obtained with full-sky coverage from a joint analysis of data collected at the Auger Observatory and the Telescope Array, where a slightly higher energy threshold was used for the Auger data ($E> 8.8$ EeV corresponding to 16,835 Auger events collected from 2004 January 1 to 2013 December 31 with zenith angle $\theta$ up to 80$^{\circ}$).  

Shown in the right-hand panel of Figure \ref{fig:8EeV} is the $D^{2}$-value distribution from 1,000,000 isotropic simulations. The reason for the increase in the number of isotropic simulations for this energy bin is the observed departure from isotropy at $\ell = 1$. The $D^{2}$-value from data, represented by the black arrow, is larger than the threshold of isotropy, presenting an indication of anisotropy with $>$ 99\% C.L.. The $p$-value obtained in this analysis from the $D^2$ estimator under the hypothesis of isotropic distribution is $ p=1.3 \times 10^{-5}$, since only 0.0013\% of the 1,000,000 simulated isotropic sky maps posses an equal or higher significance.

\subsection{Needlet analysis}
\label{sec:NeedRes}

\begin{figure}
\subfloat{\includegraphics[height=5.0cm]{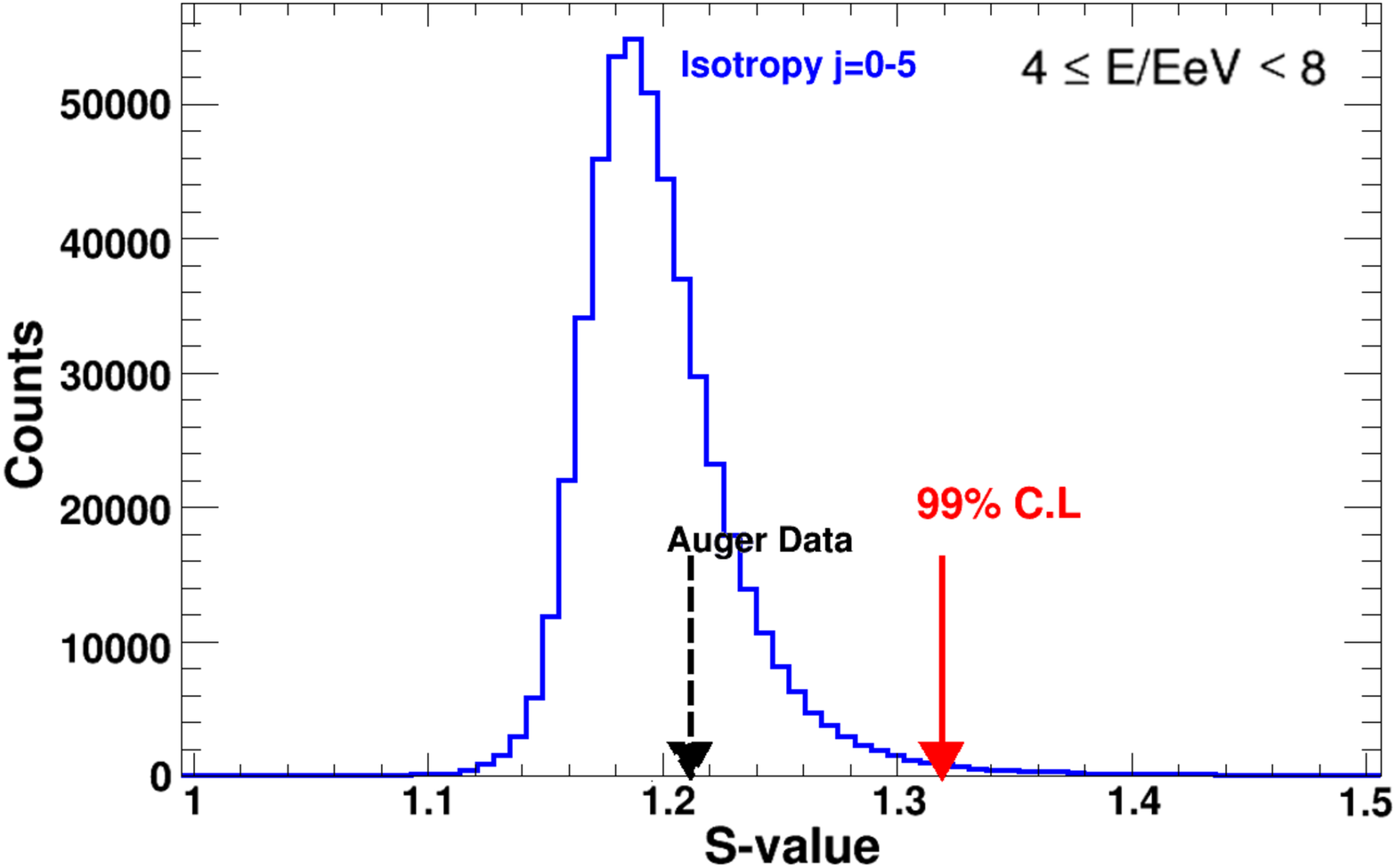}}
\quad
\subfloat{\includegraphics[height=5.0cm]{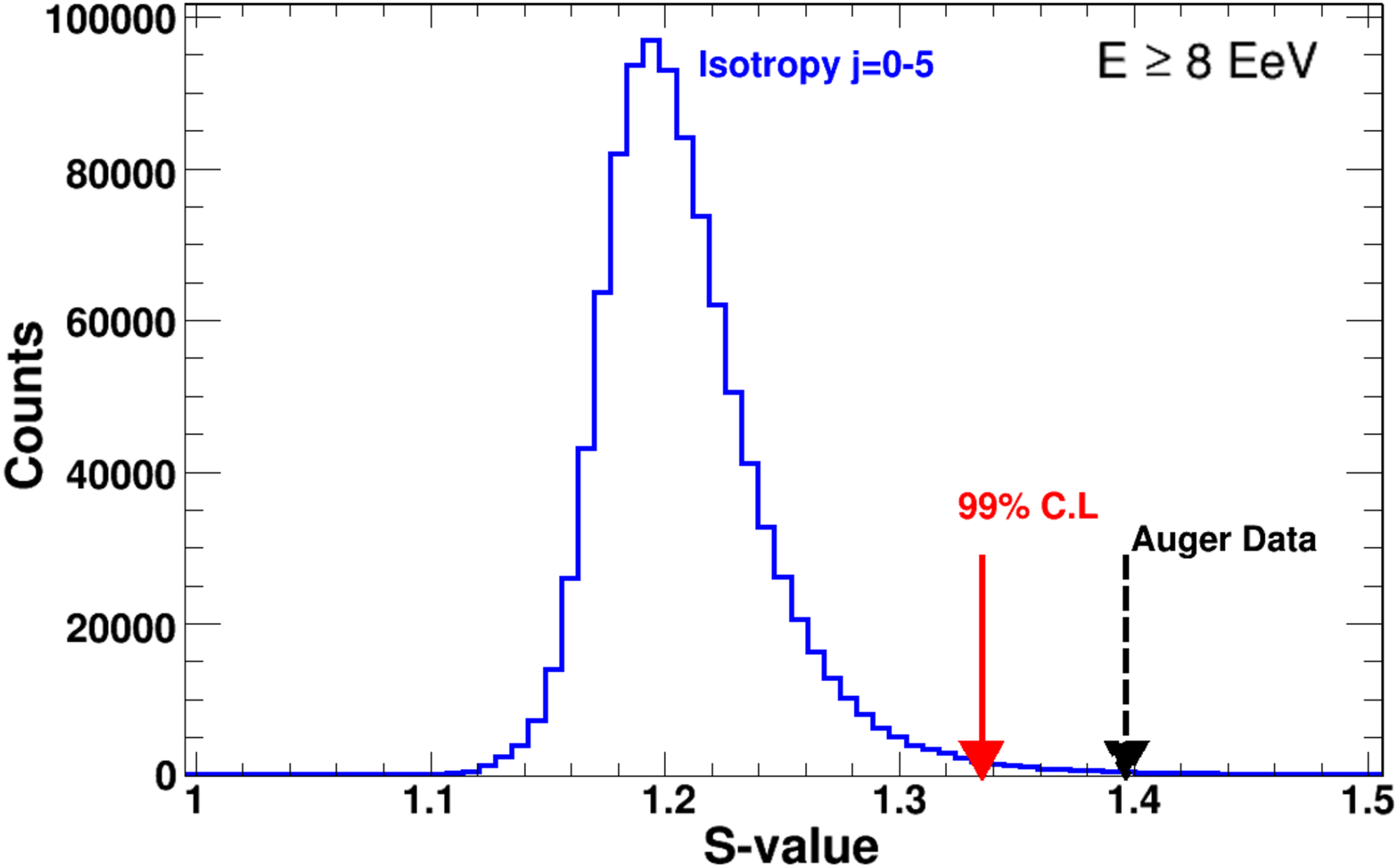}}
\caption{Results from the needlet analysis for  $j=0-5$ for  $4\leq E/\mathrm{EeV}<8$ in the left hand-panel and for $E\geq 8~$EeV in the right-hand panel. The $S$-value distribution from 500,000 (1,000,000 for $E\geq 8~$EeV) isotropic simulations for $j=0-5$ is shown.  The red straight arrow stands for the $S$-value threshold to accept/reject the isotropy hypothesis at the 99\% C.L.. The $S$-value from data is represented by the black dashed arrow. In the left hand-panel the $S$-value is smaller than that threshold supporting the isotropy hypothesis. In the right-hand panel the $S$-value from data is larger than the 99\% C.L. threshold giving an indication of deviation from the isotropy hypothesis.  }
\label{fig:4-8N}
\end{figure}

\quad As in the case of the angular power spectrum method, the data are analysed in two separate energy bins: $4\leq E/\mathrm{EeV}<8$ and $E\geq 8~$EeV. The results for the $4\leq E/\mathrm{EeV}<8$ energy bin are shown on the left-hand side of Figure~\ref{fig:4-8N}. The $S$-value distribution for $j=0-5$ of 500,000 isotropic Monte-Carlo sky maps is shown. The $S$-value of the data set is indicated by the black dashed arrow while the 99\% C.L. limit is represented by the red straight arrow. Of all isotropic sky maps 27\% have either the same or a higher significance and hence no deviation from isotropy is detected. The percentage of more or equally isotropic sky maps of the individual needlet scales is listed in Table~\ref{table_data_Needlets}.  No deviation from isotropic expectations is observed at any single scale $j$.

The results for the $E\geq 8~$EeV energy bin are shown on the right-hand side of Figure~\ref{fig:4-8N}. An indication of a deviation from isotropic expectations of the global anisotropy estimator from $j=0-5$ is observed. The probability of such a global estimator arising by chance from an isotropic distribution is $p = 2.5\times 10^{-3}$. The reason for the difference between the $p$-values obtained from both analyses is that, in accordance with Monte-Carlo studies,  the needlet method is less sensitive to large-scale anisotropies than the angular power spectrum, although it is more sensitive to small-scale patterns, not found in this study. For example, the detection power at a C.L.\ of 99\% of the angular power spectrum to a dipole with an amplitude of 7\%, located at $\delta = -40 ^\circ$ and the same number of events as in the bin with $E\ge 8$~EeV, lies around 89\% whereas the needlet method  with the same parameters used in this work achieves a detection power around 66\%. On the other hand, for a point source with a given width, the angular power spectrum can need up to twice the number of events from the source to achieve the same sensitivity.
 As pointed out in the introduction, the dipole sensitivity with the needlet method could be increased a posteriori by adjusting the parameters. This involves mainly lowering the threshold introduced in Equation~(\ref{eqn:NThreshold})  to achieve a similar or even higher dipole sensitivity than that from the angular power spectrum. The sensitivities of both methods depend on the amplitude and declination of the dipole as well as on the number of events. For instance, we mention the results concerning simulated data sets composed of 14,000 events sampled from a dipolar flux of amplitude $r = 5 \%$ and declination -30 degrees (-60 degrees).  The simulations were performed by considering the exposure of the Pierre Auger Observatory for a maximum zenith angle $\theta =  60^{\circ}$.  The detection efficiencies, at a confidence level C.L. of  $99 \%$, obtained after accounting for searches blindly performed considering all multipole moments $\ell$ up to $\ell_{max} = 64$ are $25\%$ (7$\%$) for the Angular Power Spectrum analysis and $13\%$ (5$\%$) for the needlet analysis with the same parameters used in this paper, {\it i.e.},  $B = 2$ and $T=3$. The best dipole efficiency achieved by the needlet analysis is 48$\%$ (20$\%$) for $B = 2$ and a threshold of $T=1.5$.  Again it is important to notice that this would only be beneficial in a targeted dipole search as this would severely lower the sensitivity to small scale anisotropies, going against the goal of this work: to search over a wide variety of angular scales.

Although the deviation from isotropic expectations is captured by the global anisotropy estimator, it can be observed from results for each needlet scale collected in Table~\ref{table_data_Needlets}, that only the $j=0$ needlet scale is contributing to the signal. 
\begin{table}
	\begin{center}
		\begin{tabular}{c|c|c|c|c|c|c}
			Energy range (EeV) & $j=0$ & $j=1$ & $j=2$ & $j=3$ & $j=4$ & $j=5$ \\ \hline\hline
			4 - 8 &  $51\%$ & $57\%$ & $73\%$ & $94\%$ & $57\%$ & $80\%$\\ \hline
			$\ge$ 8 & $0.0008\%$ & $58\%$ & $15\%$ & $71\%$ & $87\%$ & $83\%$ \\ 
		\end{tabular}
		\caption{ Percentage of equally or more significant isotropic sky maps  of the individual needlet scales. A look into the individual needlet scales shows that only the needlet scale $j=0$ at energies $E\geq 8~$EeV deviates from isotropy, pointing towards an anisotropy compatible with a dipolar one.}
		\label{table_data_Needlets}
	\end{center}
\end{table}

\begin{figure}
\center{}
\includegraphics[width=0.55\linewidth]{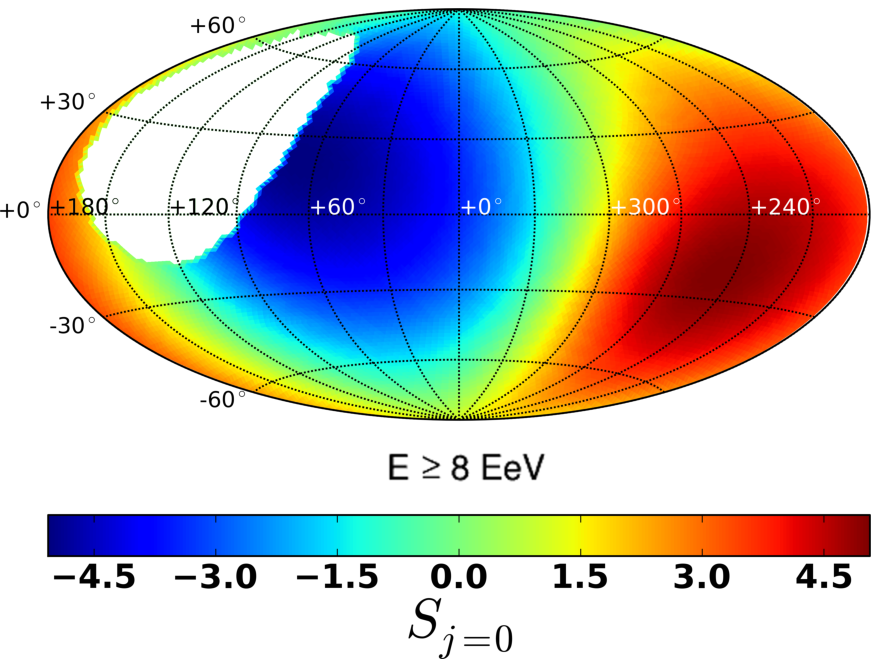}
\qquad
\includegraphics[width=0.55\linewidth]{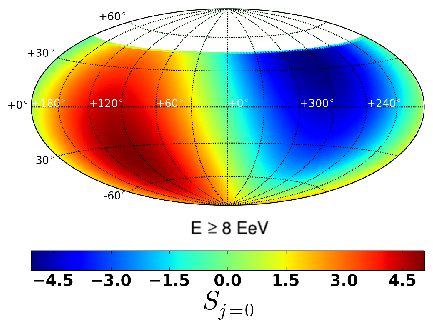}
\caption{Top part: reconstructed $S^{NT}_{j=0}$ needlet scale in Galactic coordinates for $E\ge 8$~EeV. Bottom part: the same plot in Equatorial coordinates.  
} 
\label{fig:NeedDipole8}
\end{figure}

With $B=2.0$, it turns out that the $j=0$ needlet scale corresponds to the dipolar scale $\ell=1$ in terms of spherical harmonics. For comparison purposes, and from a correspondence table built from Monte-Carlo simulations of dipolar skies, the needlet parameters can be converted into the spherical harmonics ones by adding the constraint that the dipole direction points towards the position of the pixel with the maximum significance. In this way, a dipole amplitude $d=(6.8\pm1.6)\%$ is derived pointing to a right ascension $\alpha$ and declination $\delta$ of ($\alpha,\delta) = (97^{\circ} \pm 16^{\circ},-39^{\circ} \pm 17^{\circ} )$, where the statistical uncertainties are estimated using Monte-Carlo dipolar skies with the reconstructed parameters as inputs.   These results are in  very good agreement with the reconstructed amplitude   $d=(7.3 \pm 1.5)\%$  and direction  ($\alpha,\delta) = (97^{\circ} \pm 16^{\circ},-39^{\circ} \pm 17^{\circ} )$ based on two Rayleigh analyses over the same data-set~\cite{Auger2015a} and with the joint Auger and Telescope Array analysis~\cite{AugerTAICRC2015}.  The corresponding reconstructed $S_{0k}$ sky map is shown in Figure~\ref{fig:NeedDipole8} in Galactic (top part) and Equatorial coordinates (bottom part). The reconstructed Galactic latitude $l$ and longitude $b$ of the dipole direction are $(l,b)=(247^\circ, -21^\circ)$. 

\section{Conclusions}
\label{sec:conclusion}

\quad We presented the results of two multi-resolution searches for anisotropies in the arrival directions of events detected at the Pierre Auger Observatory from January 2004 up to  December 2013 with zenith angles up to $80^\circ$. We evaluated the angular power spectrum under stationarity conditions and performed a needlet analysis in two energy bins above full detection efficiency: from 4 to 8 EeV and above 8 EeV. 

No significant departure from isotropic expectations at any angular scale is observed in the energy bin between 4 and 8 EeV for both analyses.   On the other hand, for events with energy above 8~EeV, an indication for a departure from isotropy is captured in the dipole scale, while no additional deviation from isotropy is observed at any other scale.    The dipolar amplitude and its uncertainty reconstructed by the angular power spectrum are $d = (6.0 \pm 1.5)\%$, free of contaminations from multipoles of higher order that would result from a Gaussian stochastic process shaping these multipoles in an isotropic way in ensemble average - which excludes scenarios in which the stochastic process would describe strongly constrained geometries for the source positions and produce in particular quadrupole moments tied to the observer environment.  Besides, the dipolar amplitude reconstructed by the needlet analysis is $d = (6.8 \pm 1.6) \%$, pointing in the direction ($\alpha,\delta)= (97^{\circ} \pm 16^{\circ},-39^{\circ} \pm 17^{\circ} )$. Both amplitude measurements are thus consistent with each other. The corresponding $p$-values, as obtained by the global estimators after accounting for searches blindly performed at several angular scales, are $p_{\text{APS}} = 1.3 \times 10^{-5}$ in the case of the angular power spectrum,  and $p_{\text{needlet}} = 2.5 \times 10^{-3}$ in the case of the needlet analysis. These results agree with the indications of anisotropy in the dipolar scale reported previously by the Pierre Auger Collaboration \cite{Auger2015a} using two Rayleigh analyses, and by the Auger-Telescope Array joint analysis \cite{AugerTAICRC2015}. However, as outlined in Section~\ref{sec:methods}, this is the result of a search over the whole sky observable by the Pierre Auger Observatory and does not preclude the possibility of a more localised anisotropy being detectable by a different analysis restricted to a portion of the sky. 

Future work will profit from the increased statistics, allowing us to reveal whether or not the dipole parameters derived above 8~EeV are established with larger significance. For extragalactic CRs, the firm detection of a dipole moment with an amplitude of the order as measured implies that the Milky Way is embedded into, at least, a density gradient. The eventual detection of significant multipole moments beyond the dipole one would be suggestive of significant higher derivatives than the gradient to describe the density of CRs outside from the Galaxy. Together with updated descriptions of the Galactic magnetic field to `unfold' the anisotropy parameters observed at Earth by the magnetic deflections, these future studies will help in probing both the source distribution and the propagation regime of extragalactic CRs needed to produce a density of particles outside of the Galaxy compatible with the one inferred from the kind of measurements performed in this article.

\section*{Acknowledgments}

\begin{sloppypar}
\quad The successful installation, commissioning, and operation of the Pierre Auger Observatory would not have been possible without the strong commitment and effort from the technical and administrative staff in Malarg\"ue. We are very grateful to the following agencies and organizations for financial support:
\end{sloppypar}

\begin{sloppypar}
Argentina -- Comisi\'on Nacional de Energ\'\i{}a At\'omica; Agencia Nacional de Promoci\'on Cient\'\i{}fica y Tecnol\'ogica (ANPCyT); Consejo Nacional de Investigaciones Cient\'\i{}ficas y T\'ecnicas (CONICET); Gobierno de la Provincia de Mendoza; Municipalidad de Malarg\"ue; NDM Holdings and Valle Las Le\~nas; in gratitude for their continuing cooperation over land access; Australia -- the Australian Research Council; Brazil -- Conselho Nacional de Desenvolvimento Cient\'\i{}fico e Tecnol\'ogico (CNPq); Financiadora de Estudos e Projetos (FINEP); Funda\c{c}\~ao de Amparo \`a Pesquisa do Estado de Rio de Janeiro (FAPERJ); S\~ao Paulo Research Foundation (FAPESP) Grants No.\ 2010/07359-6 and No.\ 1999/05404-3; Minist\'erio de Ci\^encia e Tecnologia (MCT); Czech Republic -- Grant No.\ MSMT CR LG15014, LO1305 and LM2015038 and the Czech Science Foundation Grant No.\ 14-17501S; France -- Centre de Calcul IN2P3/CNRS; Centre National de la Recherche Scientifique (CNRS); Conseil R\'egional Ile-de-France; D\'epartement Physique Nucl\'eaire et Corpusculaire (PNC-IN2P3/CNRS); D\'epartement Sciences de l'Univers (SDU-INSU/CNRS); Institut Lagrange de Paris (ILP) Grant No.\ LABEX ANR-10-LABX-63 within the Investissements d'Avenir Programme Grant No.\ ANR-11-IDEX-0004-02; Germany -- Bundesministerium f\"ur Bildung und Forschung (BMBF); Deutsche Forschungsgemeinschaft (DFG); Finanzministerium Baden-W\"urttemberg; Helmholtz Alliance for Astroparticle Physics (HAP); Helmholtz-Gemeinschaft Deutscher Forschungszentren (HGF); Ministerium f\"ur Innovation, Wissenschaft und Forschung des Landes Nordrhein-Westfalen; Ministerium f\"ur Wissenschaft, Forschung und Kunst des Landes Baden-W\"urttemberg; Italy -- Istituto Nazionale di Fisica Nucleare (INFN); Istituto Nazionale di Astrofisica (INAF); Ministero dell'Istruzione, dell'Universit\'a e della Ricerca (MIUR); CETEMPS Center of Excellence; Ministero degli Affari Esteri (MAE); Mexico -- Consejo Nacional de Ciencia y Tecnolog\'\i{}a (CONACYT) No.\ 167733; Universidad Nacional Aut\'onoma de M\'exico (UNAM); PAPIIT DGAPA-UNAM; The Netherlands -- Ministerie van Onderwijs, Cultuur en Wetenschap; Nederlandse Organisatie voor Wetenschappelijk Onderzoek (NWO); Stichting voor Fundamenteel Onderzoek der Materie (FOM); Poland -- National Centre for Research and Development, Grants No.\ ERA-NET-ASPERA/01/11 and No.\ ERA-NET-ASPERA/02/11; National Science Centre, Grants No.\ 2013/08/M/ST9/00322, No.\ 2013/08/M/ST9/00728 and No.\ HARMONIA 5 -- 2013/10/M/ST9/00062; Portugal -- Portuguese national funds and FEDER funds within Programa Operacional Factores de Competitividade through Funda\c{c}\~ao para a Ci\^encia e a Tecnologia (COMPETE); Romania -- Romanian Authority for Scientific Research ANCS; CNDI-UEFISCDI partnership projects Grants No.\ 20/2012 and No.194/2012 and PN 16 42 01 02; Slovenia -- Slovenian Research Agency; Spain -- Comunidad de Madrid; Fondo Europeo de Desarrollo Regional (FEDER) funds; Ministerio de Econom\'\i{}a y Competitividad; Xunta de Galicia; European Community 7th Framework Program Grant No.\ FP7-PEOPLE-2012-IEF-328826; United Kingdom -- Science and Technology Facilities Council; USA -- Department of Energy, Contracts No.\ DE-AC02-07CH11359, No.\ DE-FR02-04ER41300, No.\ DE-FG02-99ER41107 and No.\ DE-SC0011689; National Science Foundation, Grant No.\ 0450696; The Grainger Foundation; Marie Curie-IRSES/EPLANET; European Particle Physics Latin American Network; European Union 7th Framework Program, Grant No.\ PIRSES-2009-GA-246806; and UNESCO.
\end{sloppypar}

\section*{A \ \ \ Inversion of the angular power spectrum coupling matrix}

In this Appendix, we give more details on the method used in this paper to deconvolve the effects of a partial-sky coverage for recovering the underlying power spectrum. The incomplete and non-uniform exposure of the Pierre Auger Observatory gives rise to a mixing between different modes. To recover the power spectrum, the starting point is then that the harmonic transform of the product $ {\tilde\Delta}(\mathbf{n}) = \Delta(\mathbf{n}) \times W(\mathbf{n})$ is the convolution of the harmonic transforms of these functions. This way we can relate the ensemble averages of the mode-coupled angular power spectrum  $\langle{\tilde{C}_{\ell}}\rangle$ and of the true underlying one $\Braket{C_{\ell}}$ by~\cite{Hivon2002,Deligny2004}
\begin{eqnarray}
\langle \tilde{C}_{\ell} \rangle = \sum_{\ell'}M_{\ell\ell'}\left\langle {C}_{\ell'} \right\rangle ,
\label{Cl pseudo}
\end{eqnarray}
where the matrix $M_{\ell \ell'}$ is given by
\begin{eqnarray} \label{mll1}
M_{\ell\ell'}=\frac{1}{2\ell+1}\sum_{m=-\ell}^{\ell}\sum_{m'=-\ell'}^{\ell'} \int\dif\mathbf{n}_1~W(\mathbf{n}_1)Y_{\ell m}(\mathbf{n}_1)Y_{\ell' m'}(\mathbf{n}_1)\int\dif\mathbf{n}_2~W(\mathbf{n}_2)Y_{\ell m}(\mathbf{n}_2)Y_{\ell' m'}(\mathbf{n}_2).
\end{eqnarray}
Making use of the addition theorem of the spherical harmonics, $M_{\ell \ell'}$ can be expressed in terms of the Legendre polynomials as
\begin{eqnarray} \label{mll1}
M_{\ell\ell'}&=&\frac{2\ell'+1}{(4\pi)^2}\int\int\dif\mathbf{n}_1\dif\mathbf{n}_2~W(\mathbf{n}_1)W(\mathbf{n}_2)P_\ell(\cos{\gamma_{12}})P_{\ell'}(\cos{\gamma_{12}})\nonumber \\ 
&=& \frac{2\ell'+1}{2} \int_{-1}^1\dif\cos{\gamma}\int\int\frac{\dif\mathbf{n}_1\dif\mathbf{n}_2}{8\pi^2}~W(\mathbf{n}_1)W(\mathbf{n}_2)P_\ell(\cos{\gamma})P_{\ell'}(\cos{\gamma})\delta(\cos\gamma-\cos\gamma_{12})\nonumber \\
&=&\frac{2\ell'+1}{2} \int_{-1}^1\dif\cos{\gamma}~\mathcal{W}(\cos\gamma)P_\ell(\cos{\gamma})P_{\ell'}(\cos{\gamma}),
\end{eqnarray}
where $\cos{\gamma_{12}}=\mathbf{n}_1\cdot\mathbf{n}_2$, and $\mathcal{W}(\cos\gamma)$ is the two-point correlation function of $W$. 

The $M^{-1}$ matrix elements obviously satisfy $\sum_{\ell'}M_{\ell\ell'}M^{-1}_{\ell'\ell''}=\delta_{\ell\ell''}$. On inserting into this identity the explicit expression of the $M$ elements, on multiplying both sides by $(2\ell+1)P_{\ell}(\cos\gamma')/2$ and on summing over $\ell$, the completeness relation of the Legendre polynomials allows us to get at the system
\begin{eqnarray} \label{mll2}
\mathcal{W}(\cos\gamma')\sum_{\ell'}^{\ell_{\mathrm{max}}}\frac{2\ell'+1}{2}P_{\ell'}(\cos\gamma')M^{-1}_{\ell'\ell''}=\frac{2\ell''+1}{2}P_{\ell''}(\cos\gamma').
\end{eqnarray}
For a non-vanishing $\mathcal{W}(\cos\gamma')$ function, both sides of this expression can be divided by $\mathcal{W}(\cos\gamma')$. Then, on multiplying both sides by $P_\ell(\cos\gamma')$ and on integrating over $\cos\gamma'$, the orthogonality of the Legendre polynomials allows us to get at 
\begin{eqnarray} \label{mll3}
M^{-1}_{\ell\ell''}=\frac{2\ell''+1}{2}\int_{-1}^1\frac{\dif\cos\gamma'}{\mathcal{W}(\cos\gamma')}~P_{\ell}(\cos\gamma')P_{\ell''}(\cos\gamma'),
\end{eqnarray}
which corresponds to Equation~(\ref{eqn:Minv}).


\end{document}